\documentclass[12pt]{article}
\usepackage[dvips]{graphicx}

\begin{document}

\title{Variations of
parameters in nucleation process under different
external conditions}
\author{Victor Kurasov}

\date{St.Petersburg State University,
\\
Department of Computational
Physics
\\e-mail : \\
Victor\_Kurasov@yahoo.com }

\maketitle

\begin{abstract}
The nucleation process under different external conditions
is considered. It is shown that the duration of this
process can be connected with the microscopic corrections
to the free energy of the critical embryo. Connection
between variations in the value of the critical embryo free
energy and the duration of the nucleation stage is given
for several types of external conditions. This connection
is in some cases reciprocal to
uncertainty relation in quantum theory. In Appendix the
derivation of main features of the general theory on the
base of restrictions coming from the possibility of
effective and stable observations is given.
\end{abstract}

\section{Introduction}
The first order phase transitions are usually studied at
example of the transition of the supersaturated vapor into
a liquid state. This example allows to go away from the
numerous parameters characterizing the state of the mother
phase and the state of the new phase. But even in this case
there is no perfect coincidence between theoretical
predictions and results of experiments. Now it is clear
that the stationary rate of nucleation is determined with a
bad accuracy and there are serious physical reasons lying
behind this problem. So, one can speak about some
uncertainty in determination of the stationary flow of
embryos or of the stationary rate of nucleation. The bad
accuracy can be caused by two main reasons
\begin{itemize}
\item
The absence of the real stationary conditions for
nucleation.
\item
The bad value of the free energy of the critical embryo.
This value is included in the formula for the stationary
rate of nucleation.
\end{itemize}

It is clear that according to \cite{trans}
the transition of the embryos from the pre-critical zone to
the post-critical can occur far from the position of the
critical embryo and, thus, the formula for the stationary
rate of nucleation has to be reconsidered. But this case is
rather rare and here we shall consider situations when the
transition from the pre-critical zone to the post-critical
zone goes through the critical point, i.e. through the
position of the critical embryo.

The numerous investigations of the establishing of the
stationary state in the near-critical region showed that
the situation where the stationary state is not established
are very rare also. To see such situations one has to cut
off the power of metatstability practically immediately
after the moment when this metastability was created. So,
this opportunity is also out of consideration here.

The paper is organized as following: In the next section
the situation with the artificial cut-off of the
supersaturation is analyzed. Then the situation of decay of
metastable phase is considered. Here the form of relation connecting
the variations of parameters of the process resembles the
uncertainty relations in quantum mechanics. That's why in appendix
the derivation of the basic characteristics of the calssical
mechanics and quantum theory including the uncertaity
relations is given.
The last situation which is considered is the external
conditions of graduate creation of metastability in the
system. Here the form of relations connecting the same
parameters of the process radically differs from those in
the
previous situation.

\section{Determination of the pure rate of nucleation}

The stationary rate of nucleation in the main order has
rather transparent origin - the rate of nucleation is
proportional to exponent of the free energy $G$ of the critical
energy taken in thermal units. Here and later all
 values having the sense of energy are taken in thermal
 units $kT$, where $k$ is the Bolztman's constant and $T$
 is the absolute temperature.

Determination of the free energy $G$ is a very complex
procedure. The problem is that the critical embryo has a
number of molecules $\nu_c$ big enough to make useless all
calculations based on dynamic laws of motion. On the other
hand the number $\nu_c$ is not big enough to apply the laws
of statistical mechanics. But since there is no alternative
one has to use the approach of thermodynamic description.

Unfortunately the situation is more dramatic because one
has to calculate the exponent of the free energy. Although
the relative error in determination of the free energy
becomes small the exponent reflects the absolute errors
and these errors are not small even with $\nu_c \rightarrow
\infty$.

Really, the extraction of a separate
embryo from the whole system is some
simplification. It works satisfactory because
the intensity of exchange between the embryo and
environment is much more weak than the intensity of
relaxation in the
embryo to the state of internal equilibrium
$$
t_{int} \ll t_{ch}
$$
where $t_{int}$ is the time of internal relaxation and
$t_{ch}$ is the characteristic time of exchange with
environment.
But here appears a problem -
it is impossible to determine concretely the
type of conservation equations for the separation of the
embryo. So, it is impossible to determine the ensemble in
statistical mechanics.

From the first point of view there is no problem because
different  ensembles in statistical mechanics give
equivalent results. But this means only that the relative
values of macroscopic variables are equivalent. More
precisely the relative difference of values have the order of
$\ln n/n$ where $n$ is a number of particles in the system.
Taking exponent one can see the difference in $n$ times.

Also one has to take into account the possibility of
fluctuations which gives for the thermodynamic
potential the shift of the order $n^{1/2}/n$
i.e. $n^{-1/2}$. So, the exponent will have the correction
in $\exp(\sim n^{1/2})$ times. This correction is
enormously big. Certainly, one can say that fluctuations
are already taken into account but the trace of
incompleteness of the theoretical derivation still remains.

The next source of inevitable difficulties is the
limitations on the size of the embryo. There are two
aspects of this problem
\begin{itemize}
\item
Since the system has a finite dimension the continuous
spectrum of the energy transforms into the discrete energy
levels and all integrations have to be replaced by
summations. It is vary hard to do because the
Euler-Maclaurin's
decomposition is not converging  one, but only asymptotic
and the formal  transformation here is not
possible.
\item
There appear the simple geometric problems like the
difference between position of the  surface of tension and
the equimolecular surface. The conception of the surface of
the embryo which is necessary to write the term with the
surface energy and the equation of the material balance.
But positions of surfaces do not coincide. This leads to the
additional term of the order of $n^{1/3}$.
\end{itemize}

Following the second item we shall write the formula
for the free energy as
$$
G = - b\nu + a \nu^{2/3} + c \nu^{1/3}
$$
where parameters $a$, $b$, $c$ have a simple physical
meaning: $b$ is the difference of chemical potentials in a
mother and a new phase, $a$ is the renormalized surface
tension, $c$ is connected with the difference between
equimolecular surface and the surface of tension at the
plane surface. Later there appear corrections due to the
curvature of the  surface and it is convenient to continue
this decomposition writing it as
$$
G = - b\nu + a \nu^{2/3} + \sum_{i=-1}^\infty c_i \nu^{-i/3}
$$
The term with $i=0$ looks like $const + c_0 \ln \nu$ and it
is connected with non-equivalence of ensembles.

As the result of all these constructions one can state that
there is a microscopic addition $\delta G$ to the value of
$G_0$, there is also an addition $\delta \nu_c$
to the argument of maximum
$\nu_{c0}$. Here the values with a subscript $0$ are the
values based  on the capillary approach, i.e. on
$$
G = - b\nu + a \nu^{2/3}
$$

The experiments intended to get the rate of nucleation and
the free energy of the critical embryo are ordinary
constructed in a following manner:
At the initial moment of time there is a metastable state
with no embryos of the new phase.
After some time $t_{cut}$ the metastability in
the system is artificially diminished to forbid the
formation of new embryos. Then the number of droplets will
be
$$
N = J t_{cut}
$$
Here no depletion of the vapor phase is taken into account.
To neglect the depletion of the mother phase it is necessary
to fulfill
$$
t_{cut} < t_{depl}
$$
where $t_{depl}$ is the time of depletion, which will be
determined in the next section.
Then
here
$$
\delta t =0
$$
and
$$
\delta N = (\exp(\delta G)-1) N_0
$$
For very small $\delta G$ one can linearize the exponent and
get
$$
\delta N = \delta G N_0
$$

\section{Decay of metastable state}

Now we shall consider the process of the mother phase
depletion. To give quantitative estimates it is necessary
to specify the rate of the droplets growth. For the
supercritical embryos, i.e. for the droplets it is
reasonable to adopt the free molecular regime of growth.
Under this regime the question of profile of the mother
phase around the droplet can be solved extremely simple -
there is no such a profile and, thus, the mother phase
depletion can be described in a very simple manner.

Under the free molecular regime of growth the number of
molecules inside the droplet grows as
$$
\frac{d \nu^{1/3}}{dt} = \zeta/\tau
$$
where $\zeta$ is the supersaturation of the mother phase
and $\tau$ is some characteristic time which is
approximately a constant value.

Then the number of molecules inside the new phase will be
$$
Q= \int_0^t J(t') \nu(t') dt'
$$
where $\nu(t')$ is the number of molecules inside the
droplet formed at $t'$.

The rate of  nucleation is connected with the distribution
function $f$ over $\rho=\nu^{1/3}$ by
$$
f=J\tau/\zeta
$$
Then
$$
Q= \int_0^z f(z) (z-x)^3 dx
$$
One can show that the depletion occurs in a very rapid
avalanche manner.
Before the essential depletion one can  consider $f$ as a
constant $f_*$ at the beginning of the process and get
$$
Q = f_* z^4/4
$$
The length of the spectrum is determined by the following
condition
$$
Q = \frac{\zeta}{\Gamma}
$$
where
$$
\Gamma = - \zeta \frac{dG(\nu_c)}{d\zeta}
$$
Then
$$
f_* z^4 = \frac{4 \zeta}{\Gamma}
$$
and one can determine the time $t_{depl}$ from the
following condition
$$
J_* \frac{\tau}{\zeta} t_{depl}^4 (\frac{\zeta}{\tau})^4  =
\frac{4 \zeta}{\Gamma}
$$
where
$$
J_* = f_* \zeta/ \tau
$$
Having written $J$ as $Z\exp(-G(\nu_c))$ where $Z$ is
the Zel'dovich' factor one can come to
$$
Z \exp(-G)  t_{depl}^4 (\frac{\zeta}{\tau})^3  =
\frac{4 \zeta}{\Gamma}
$$
The last equation allows the analysis of variations of the time
on depletion and the free energy of the critical  embryo
formation.

Having inverted variations one can get
$$
Z \exp(-G-\delta G)  (t_{depl}+\delta t)^3
(\frac{\zeta}{\tau})^4  =
\frac{4 \zeta}{\Gamma}
$$

This is the final equation and one can see that the
total number of droplets can be calculated as
$$
N = J_* t_{depl}
$$
In the main order
$$
N \sim \exp(- 3G/4)
$$
and
$$
\delta N \sim N_0 (\exp(- 3\delta G/4)-1)
$$

The equation on $\delta G$, $\delta t$ can be linearized
which gives
$$
\frac{\delta t}{\delta G} = \frac{t_{depl}}{4}
$$
This equation is reciprocal in its functional form to the
uncertainty relation in quantum mechanics
$$
\delta E \delta t = const
$$
which combines the uncertainty in energy $E$ and in time
$t$. That's why  in Appendix the method based on
uncertainty relation is presented. It is necessary to
stress that the derivation in appendix has a special
meaning and demonstrates some new features.

\section{Gradual creation of a supersaturation}

Ordinary the external conditions have a continuous slowly
varying character. Then one can introduce the ideal
supersaturation $\Phi$, i.e. a supersaturation which would be in
the system in the absence of formation of a new phase. Thus,
 the ideal supersaturation is fully governed by external
 conditions. At the variations of $\Phi$ of a relative
 order of $\Gamma^{-1}$
$$
\delta \Phi = \Gamma^{-1} \Phi
$$
the behavior of $\Phi$ can be linearized
$$
\Phi(t) = \Phi_* + \frac{d\Phi}{dt} (t-t_*)
$$
Here $*$ marks values at some characteristic moment.

The duration of the nucleation period is approximately $2t_{nuc}$
where $t_{nuc}$ satisfies the evident relation
$$
t_{nuc} \frac{d\Phi}{dt}|_* = \zeta_* \Gamma^{-1}_*
$$
So, the value of $t_{nuc}$ is absolutely independent on
$\delta G$ and, thus,
$$
\delta t_{nuc} =0
$$

The moment $t_*$ has to be chosen as the moment of the
maximal intensity of the droplets formation, i.e. here as
th moment of the maximal supersaturation. Then here
$$
\frac{d\zeta}{dt} =0
$$
and
$$
\frac{dQ}{dt} = \frac{d\Phi}{dt}
$$

The variation $\delta G$ certainly exists, but does it take
place the variation of $G_*$? To see this variation one has
to write the condition for the maximum of the
supersaturation
$$
\int_{-\infty}^{t_*}
J_* \exp(\Gamma \frac{d\Phi}{dt}|_* (t'-t_*))
\frac{\tau}{\zeta} \rho(t')^2 \frac{3\zeta}{\tau} =
\frac{d\Phi}{dt}|_*
$$
One can get $J_*$ outside of the integral and see that
the previous equation reduces to
$$
J_* = slow\ \ function \approx const
$$
It means that $J_*$ is invariant. When
we add to $G$ some addition $\delta G$ nothing will be
changed. Simply the moment $t_*$ will be attained earlier
or later. Then it is possible to  introduce a shift of $t_*$
and this shift will depend on $\delta G$.

As the result one can state that in this case there is
no variations of parameters $\delta G_*$, $\delta t$.

\section{Conclusion}

Having analyzed three concrete situations one can see that
only in the situation of decay there is a variations of
parameters and these parameters forms the relation
reciprocal to uncertainty relation in quantum theory.

\appendix

\section{The role of restrictions coming
from the possibility of stable
and effective calculations  }

At first the aim of this review was to show the role of
requirements coming from the possibility of correct
calculations in physics. Later the constructions based on
the requirement to produce the stable calculations of the
characteristics of the system gave some more general
conclusions presented below.

We start from the general point of view and instead of
concrete physical theory we consider the general
qualitative causal theory based on the differential formalism.
It is necessary to clarify the last sentence. When we
mention
"the differential formalism" it means that the theory uses
the standard formalism of the differential calculation. The
term "causal" shows that some events are considered as
"reasons" of other events described as "sequences".

There is no other special assumptions to start our
constructions, but later some rather evident notations will
be made to give us the possibility to present concrete
results.

\subsection{Causality and time}

It is necessary to introduce a variable (or a
characteristic) to describe the property of causality.
Really, our style of thinking is principally a causal
style. But the qualitative formalism has to show what event
is a reason and what event is a consequence. In everyday
speech we use the terms "earlier" for the reasons and
"later" for the consequences. For example, consider the
number of cars and the number of crushes. It is clear and
mathematical statistics can show  that the number of
crushes correlates with the number of cars. But what is the
reason - the big number of cars or the big number of
collisions? Mathematical statistics can not give us the
answer. It attracts our attention only to the fact the the
increase of cars is associated with the increase of
collisions. Certainly, we know that the increase of the
cars is the reason of collisions (However, collisions are
in some sense the source of beaten cars and leads to
increase of the total amount of cars, but beaten cars are
excluded from consideration). But how it can be proven?
Only by the fact the the increase of the number of cars
occurs slightly  earlier than the increase of
the number of collisions. In
real macroscopic
social systems one can not really observe this effect
obviously and this produces additional difficulties.
But our style of thinking is to search the reason and the
reason is marked by the word "earlier".

It is quite possible to see the system where the number of
collisions is the reason of the number of cars. Really
consider the social or biological system where "collisions"
are "sexual relations" and the role of cars is played by
males and females. Then the opposite casual construction
takes place. The number of collisions is the source of the
number of males and females. And here one can say that
collisions occurs earlier than the increase of the population.

The straight result of the given example is the necessity
to introduce the the characteristic responsible for the
casuality. This characteristic is called as the "time". It
will be marked by a letter $t$. The task of the theory which is going
to be constructed is to determine the dependence of characteristic
of the system $x$ on $t$, i.e. $x(t)$. If there are several
characteristics $x_{(i)}$
of the system which are marked by  the index
$i$ one can consider a vector $\vec{x}$. At first we shall
consider the case of one variable $x$.

One can argue whether $t$ is discrete or continuous. The
arguments to consider discrete time $t$ are connected
with a quantum Zenon effect. To use advantages of the differential
formalism we consider here $t$ as continuous variable, at
least at some first steps of our considerations. The
interval for time will be $[a,b]$. Sometimes we shall take
it
as an interval $[-1,1]$.

Here we have to state that the reversibility ordinary
announced in the classical mechanics has absolutely another
rather local sense. To see the formal reversibility one has
to change all velocities and the there is no concrete
method how to do this. So, this reversibility is only
imaginary property. In a real world there is no way to
change the direction of time. Time is the characteristic
responsible for the causal relations in our world.

\subsection{Properties of proximity}

The function $x(t)$ has to be established by the theory and
then it has to be checked by some experimental measurements.
The measurement of $x$ at the moment $t_i$ will be marked
$x_i$. Certainly, there is a characteristic error $\delta
t$ of the choice of the time moment. At the accurate
measurements this error becomes infinitely small. Then to have an
infinitely  small error of $x$ it is necessary that $x(t)$
has to be a continuous function
$$
x(t) \in C_{[a,b]}
$$

Here we have to choose the measure of proximity. According
to the central limit theorem of the probability theory
the distribution of errors of stochastic variable
(let it be $y$)  under some rather
wide spread conditions goes to the normal
distribution $N$
$$
N \sim \exp( - \alpha^2 (y - \bar{y})^2)
$$
where $\bar{y}$ is the mean value of $y$ and $\alpha$ is some
constant.

The last relation leads to the choice as the most
appropriate metrics the ordinary  metrics
$$
|| \vec{x} || = \sqrt{ \sum_{i=1}^n x_{(i)}^2 }
$$
This metrics corresponds to the  scalar product
$$
<x,y> = \sum_{i=1}^n x_{(i)} y_{(i)}
$$

The last scalar product corresponds to the Pyphagorean theorem
for orthogonal basis
$$
\alpha^2_i x_{(i)}^2 + \alpha^2_j x_{(j)}^2 = x_{(ij)}^2
$$
where
$$
\alpha_i \vec{x_{(i)}} + \alpha_j \vec{x_{(j)}} = \vec{x_{(ij)}}
$$
is treated as a vector sum and $\alpha_i$, $\alpha_i$ are
some constants.
It is necessary to stress that $\vec{x_{(ij)}}$ will be
orthogonal to all $\vec{x_{(k)}}$ with $k \neq i,j$ and the
given property can be used again and again.

The necessity to use this property for our construction
is the following. We have to stress that we do not know the
"true" characteristics of the system. We can miss some of
them. There is possible to see the situation
when instead of a pair coordinates we take one coordinate
which is a linear combination of the initial ones. But the
form of the normal distribution $N$ has to be the same as it stated
by the central limit theorem. It is
possible only when we take the mentioned scalar product.

Really, for the probability of two independent
characteristics $x_{(i)}$ and  $x_{(j)}$ we have
$$
P = P_i P_j = \exp( - \alpha_i^2 x_{(i)}^2)
\exp( - \alpha_j^2 x_{(j)}^2)
= \exp( -  x_{(ij)}^2)
$$
Here we count $x_{(i)}$, $x_{(j)}$ from their mean values.

\subsection{Dimensionality of a physical space}

On the base of measurements we have to reconstruct the
function $x(t)$. Consider the simplest case which is the
case of one material point, i.e. a simplest system without
any external ad internal parameters and characteristics.
How many coordinates is necessary to introduce in order to
describe this system?
The evident answer is that the simplest case is one
coordinate. But this answer has one disadvantage which will
be considered below.

Suppose that $x(t)$ is some signal which is governed by
stochastic process of random motion.
The results of Poia \cite{Lamperti} show
that for the stochastic walking the return back to origin
with the probability $1$ will be infinitely many times when
the dimension of space $d$ is $d=1$ or $d=2$. When $d=3$ or
greater then the probability of the infinite number of
returns is $0$.

The illustration of these results can be easily seen if we
mention that the diffusion equation
corresponding to this process has the Green function
 with essential part
$$
G \sim \exp(- \sum_i x_{(i)}^2/4D_{(i)}t)
$$
where $D_{(i)}$ is corresponding diffusion coefficient.
The rest in $G$ is the normalizing factor depending only on
$t$.

We see that the functional
form of $G$ does not depend on the number of the spatial
variables. Here lies one of the possible reasons why we see
the diffusion process clearly.
This functional form coincides with the
functional form of the normal distribution. Again this form
is the exponential of the square form of the variables.
This allows to speak about the distributions of this form
as the result of the random walking process and the
fundamental functional form which will be used below.
Again one can see the invariant character towards the
choice of the variables or their ignorance.

Now we have to describe the consequences of these results
for the problem under consideration. When $d=1;2$ the
infinite number of returns allows to construct the infinite
set of measurements at the moments of these returns and
have all measurements as the zero values. So, the trace of
the random walks disappears. This effect is unsatisfactory
and we need to have $d=3$ or greater to exclude this
effect.

Since there is no other characteristics of the material point
there is no other candidates for the true dimension of the space
and we have to admit that namely $d=3$ is the crucial
dimension.

\subsection{Reconstruction of functional dependencies}

The number of measurements of trajectory $\vec{x}(t)$ is a
big finite number going to infinity. On the base of these
measurements one has to reconstruct the functional form
$x(t)$ (here it is sufficient to consider one variable).
Since there are only two arithmetical operations  (addition
and multiplication) one can not go outside polynomials.
Actually only polynomials can be constructed and
calculated.
All other functions which are ordinary used like $\sin$,
$\cos$ are no more than idealized infinite series of
polynomials.

We have to restrict the class of functional dependencies
allowed for $x(t)$. Really, for discontinuous functions one
has to measure $x(t)$ at every point $t$ which is certainly
impossible. So, the consideration of the class of
continuous functions is preferable not only from the
physical point of view but also from the  enormous
expenditures of measuring.
Fortunately according to the
Weierstrass theorem every continuous function $f$ at $[a,b]$ can be
approximated by polynomial $P$
$$
|| f-P ||_{C_{[a,b]}} < \epsilon \rightarrow 0
$$
or
$$
max_{a<x<b} | f-P | < \epsilon \rightarrow 0
$$

The simplest form of approximation is interpolation.
The property of interpolation means that having measured
 $n$ times at moments $t_i, i=1..n$
 the function $x(t)$ we get $x_i = x(t_i)$ and the
 polynomial $L$ has a property $L(t_i) = x_i$.
Certainly, it is possible to construct the polynomial of a
power $n-1$ in a unique form.

But
here one faces with the "no go theorem" which states that for
every manner of the choice of  interpolation nodes there
exists a function which can not be interpolated
\cite{Lebedev}. Namely, the difference between function $f$
and interpolation polynomial $L$ has the estimate
$$
|| f-L ||_{C} = \it{O} (\ln n)
$$
where $n$ is the number of nodes.



 There exist recipes of Feier and Valle-Poussin
\cite{Lebedev} which allow to approximate $f$ in $C$, but
these recipes have  no property of interpolation: there
exist nodes where $P(t_i) \neq x_i$. So, we came to a
strange situation with a trajectory which does not satisfy
the results of measurements. Some analogy takes place in
quantum mechanics.

Having started from the classical
point of view we have to require that all measurements have
to be satisfied precisely.

It is not a accidental coincidence that in the method of
Valle-Poussin the approximated function does not coincide
with the measurements at the half of points. This number
can be hardly decreased because the weight used in this
method is the optimal choice \cite{Lebedev}.

We come to a strange conclusion that at  least
some measurements can not precisely define the investigated
dependence. This can be explained by impossibility of the
fully precise determination in experiments of all possible
characteristics of the system. In quantum mechanics this
effect is called as "uncertainty relations".

\subsection{Restriction of possible trajectories}

The possible evident  answer to solve the problems
leading to a classical mechanics appeared from
the "no go" theorem is to consider instead of the space
$C_{[a,b]}$ the space of functions with restricted first
derivative. For such functions one can see \cite{Bahvalov}
that for every $t$
$$
| x(t) - L(t) |  \leq  \it{O} (\ln n / n)
$$
where $n$ is the number of nodes (the number of
measurements). Here the nodes are the Chebyshev's ones. The
interval is $[-1,1]$.
So, here the interpolation procedure approximates the real
trajectory. Now the (infinite) set
of  measurements can give us the form of
trajectory.

There appeared two important consequences:
\begin{itemize}
\item
There appeared a new auxiliary  characteristic of the system
- the velocity $v$ or the momentum $p$. Now the description
has to take into account this characteristic explicitly.
\item
The velocity of trajectory is limited by some constant
$c$. This corresponds to the requirement of the special
theory of relativity. So, one can assume that the special
theory of relativity goes from this very simple restriction
of the  class of trajectories in order to have the
convergence in procedure of interpolation.
\end{itemize}

As the special result we come to a conclusion
that the trajectory and the first derivative of trajectory
are the basic characteristics in description of the state
of the system (or of the particle).

One has to stress that $p$ or $v$ can not be considered as
the variables fully equivalent to coordinate $x$. In
classical mechanics there exists a picture of Hamilton
where $x$ and $p$ are formally considered
as a pair of coordinates. But one has to remember that
initially $x$ and $p$ have different senses and $x$ is
the main variable, while $p$ is additional one. In
consideration presented here it
appeared as the characteristic only because
the restriction of the class of trajectories.


\subsection{The configuration space and the phase space}

At first it is necessary to
recall that there exists a simple style to present
the state of a complex system - the configuration space and
the phase space.

The configuration space is a space $R^n$ where $n$ is a
number of coordinates of all particles in a system. The
state of the system is a point in configuration space. The
coordinates in configuration space are orthogonal and this
is one of essential features in the future analysis. This
orthogonality is an evident consequence of the simple fact
that if different particles in the system are independent
then the space is simply reduced to the sum of two
configuration subspaces for particles (Certainly, the
configuration space for a free material point is $R^3$).
We assume this fact at least for negligibly interacting
parts of the complex system.

When we add the momenta as auxiliary characteristics of
the system we get the space $R^{2n}$ of all coordinates and
momenta of the  system. Here also orthogonality takes
place. One can say that this is simply the property of a
linear orthogonal space or one can seek something behind
this fact.

The problem which appears for every theory is that one does
not know the number of coordinates of the system. The
simplest
structure of the physical construction implies that we have
a system with the given number of interacting balls. They
are referred as "particles". Certainly, one can not state
that these particles are the simplest systems, they also
have to be considered as complexes with rather complicated
structure. In the field models the number of degrees of
freedom is principally unknown. So, our approach has to
allow the generalization to unknown coordinates, it has to manifest
some invariance for the squeezing and for the developing of
description.
In approach to construct the field theories by continual
integration procedure it is also necessary that the kernel
has to be invariant for the arbitrary choice of the number
of coordinates \cite{Fein}.

The subspace responsible for internal degrees of freedom
has to be separated in some sense from the subspace
responsible for external degrees of freedom. The best way
to ensure this property is to consider these subspaces as
orthogonal ones. The property of orthogonality is
associated with the Pythagorean  theorem
$$
c^2 = a^2 + b^2
$$
and with Euclidean postulate that at the given point
one can put only one
parallel line to the given line. This postulate leads to
precise recipe resulting in the unique geometric operation.
Then this postulate can be considered as reflection in the
everyday life of intention to construct
quantitative theory describing the world.

What functions satisfy the requirements put here?
It is easy to see that only the square forms of
characteristics (coordinates) allow the operation of
squeezing. Really, for
$$
F = \sum  a_{(i)} x_{(i)}^2
$$
with some constants $a_{(i)}$
if we miss some $x_{(i)}$ the functional form remain the same
and is instead of $x_{(j)}, x_{(k)}$ we take a linear
combination $x_{l} =
b_j x_{(j)}+ b_k x_{(k)}$ the form will be the invariant
also.

Certainly, the square form can be reduced to the sums of
squares by a linear transformation.

\subsection{Classical motion}

The knowledge of the function $x(t)$ means that we know the
functional form for $x(t)$ which can be written as
$$
H(x(\tilde{t}),t) = const
$$
for some function $H$, where $\tilde{t}$ means that all preceding
times are involved in description.
The function $H$ is called an "energy" or having written as
a formula it is called the "hamiltonian".
Giving more detailed  description at the current
moment $t$ we can indicate the derivatives of trajectory
$$
H(x,dx/dt, d^2 x/ dt^2, ...,  t) = const
$$
Since we have
established that only the value $x(t)$ and
the value of the first derivative $dx(t)/dt$ are included into
description then we have
$$
H(x,dx/dt,t) = const
$$

The explicit dependence on time has to be excluded since
all behavior of the system has to be invariant in respect to
$t-t_0$ where $t_0$ is the time of preparation of initial
state.
Then we get
$$
H(x,dx/dt) = const
$$

For the dependence on $v=dx/dt$ since as it has been
mentioned $v$ is a formal auxiliary characteristic we have
to take the formal square functional dependence as it is
prescribed above
$$
H(x,v) = U(x)+ m v^2/2
$$
where $U$ is some function and $m/2$ plays the role of
$a_{(i)}$

The requirement $H=const$ or $dH/dt=0$ leads to
$$
mdv/dt=-\frac{\partial U}{\partial x}
$$
which the Newton's second law of motion.
The trajectory of motion $x(t)$ will be called the classical
trajectory.

The same transformations can be done for vectors $\vec{x}$
and $\vec{v}=d\vec{x}/dt$.

The function $-\frac{\partial U}{\partial x}$ is ordinary
described as the sum of forces. Fortunately this function
has a rather simple form.

\subsection{Statistical mechanics}

We have marked that the complete set of characteristics of
the system can not be presented even for the simple
systems.
The account of the missed degrees of freedom or missed
parameters can be made in a manner
like it is done in statistical mechanics. Here one can use
the derivation presented in \cite{Landau-stat}.
Instead of precise characteristics
$x_{i)}, v_{(i)}$ one has to use the distribution function
$\rho$
which is the probability $P$ to find characteristic $q$ in the
interval $[q,q+dq]$ divided by $dq$
$$
\rho(q)  dq = P(q \in [q,q+dq])
$$
When for
$H$ one can see
$$
H=H_1 + H_2
$$
where $H_1$ is the hamiltonian of the first group of
particles and $H_2$ is the hamiltonian of the second group
of particles then two subsystems are independent.
It is shown directly from dynamic equation of motion.  Then the
distribution function
$\rho(H)$ has to be the product $\rho(H_1) \rho(H_2)$
$$
\rho(H_1+H_2) = \rho(H_1) \rho(H_2)
$$
The last functional equation has an evident solution
$$
\rho \sim \exp( \beta H)
$$
which determines the distribution function with a parameter
$\beta$, which is proportional to inverse temperature.

Certainly, in foundations of the statistical mechanics it
is necessary to see the hypothesis of Boltzman which states
that the averaging over trajectory is equal to the averaging
over ensemble. Rigorously speaking one has to prove this
fact based on dynamic equations but it is very difficult to
do.

Since the distribution has to be normalized
$$
\int d \Gamma \rho(H) = 1
$$
where $\int d\Gamma$ is the integration over all possible
states,
one can determine the normalizing constant $b$ in
expression
$$
\rho = b \exp(-\beta H)
$$
Then
$$
b = [\int d\Gamma \exp(-\beta H)]^{-1}
$$

One has to show that the missed coordinates do not change
the functional form of the distribution function. Really,
for $H=H_1+ a q^2$ we have
$$\rho (H) = \rho(H_1) \exp(-\beta  a x^2)$$
and $\rho(H)$ differs from $\rho(H_1)$ only by the factor
which coincides in the functional form with a normalizing
factor.

The same derivations can be  done for
the arbitrary choice $x_{l} =
b_j x_{(j)}+ b_k x_{(k)}$
and the factors of the normal distribution appear.

The coordinates $v_{(i)}$ can be separated
due to the form of $H$. This leads to the Maxwell
distribution over velocities \cite{Landau-stat}
$$
\rho(v_{(i)})
\sim
\exp(-m v^2_{(i)} / 2)
$$

Summarizing this section we introduce the recipe to take
into account the missed degrees of freedom by transition
\begin{itemize}
\item
$$ H \rightarrow \exp(-\beta H) $$
\end{itemize}

\subsection{Simple solutions}

The linear dependence of $H$ does not lead to any progress.
Now we shall present the solutions initiated by the
simplest square
form of $H$, i.e. $H=ax^2 + b v^2$.
From the last section to have the convergence
we see that constants $a$ and $b$
have to has the same sigh as $const$ in $H=const$. We have
three possibilities
\begin{itemize}
\item
$a=0$

Then $v=const$ and we
have the straight uniform motion which is equivalent to
the stationary state.

\item
$b=0$

Then $x=const$ and there is no motion. This case does not
take place

\item
$a>0, b>0, const >0$

The solution is
$$
x = A \sin w t + B \cos w t
$$
where $w$ is the frequency of oscillations.
\end{itemize}

The last case is the most common one. If instead of $x_i$
we choose $x_{i+1} - x_i$, where $x_i$ are coordinates
of particles  we get the moving wave which
satisfies the wave equation
$$
\frac{\partial^2 f}{\partial t^2}
= \kappa \frac{\partial^2 f}{\partial x^2}
$$
for a function $f$
with a positive parameter $\kappa$.

\subsection{Uncertainty relations}

The base of the classical mechanics is in some sense
contradictory. The value of velocity involved into
the formalism of the classical mechanics is an ill defined value
from the point of view of numerical methods. Really, the
numerical definition of derivative
$$
v  = \lim_{\delta t \rightarrow 0} \frac{x(t+\delta t ) -
x(t) }{\delta t}
$$
has  a problems in calculation $ \frac{x(t+\delta t ) -
x(t) }{\delta t}$ for small $\delta t$.

From a physical point of view of classical approach one can
see that collisions with environment make the velocity of
the Brownian particle a fluctuating value and
the instant value of velocity can not be well determined
while the coordinate of the Brownian particle is a stable
characteristic.

These features require to consider some characteristic
values of fluctuations
$\delta x$ and $\delta v$ of
$x$ and $v$.
The concrete form of uncertainty relation can be
established on the base of quantum mechanics which is
derived in next sections. Here one can stop the narration
of this section.

These simple features show us  that it is necessary to
present some approach which has to take into account the
impossibility to have in one and the same moment the value
of $x$ and $v$.

But also one can present some not so rigorous derivations
leading to some interesting results. already the
Liouville's theorem says that the volume in the phase space
is conserved and, thus, $\delta p \delta x$ which is the
elementary volume is conserved.

Analogous derivations following Hazen start from
the Hamiltonian form of the law of motion
$$
\frac{\partial H}{\partial p} = x'\ , \ \
\frac{\partial H}{\partial x} =  - p'\
$$
where a sign $'$ marks the derivative on time
and an evident
Maxwell relation
$
\frac{\partial^2 H}{\partial p \partial x} =
\frac{\partial^2 H}{\partial x \partial p}
$
and one can see that
$
\frac{\partial x'}{\partial x}
+
\frac{\partial p'}{\partial p}
= 0
$.
Since for every function $f$
$
\frac{\partial f'}{\partial f } =
\frac{f''}{f'}
$
one can
find
$
p'x''+x'p''=0
$
or
$
\frac{d}{dt}(p'x') = 0
$.
In terms of finite differences
$
\frac{d}{dt}(\delta p\delta x) = 0
$
states that $\delta p\delta x$ remains some
constant.

Now one can consider relation $\delta E \delta t$.
certainly, the system following dynamic equations has
$\delta E =0$. Then one can not use here Hamiltonian
relation but simply consider $E=E(x(t))$. Then
$dE=\frac{dE}{dx} \frac{dx}{dt} dt$. Since
$\frac{\partial E}{\partial x} = p'$ which can be regarded
as a definition of momentum. Then $dE=p'x'dt$. Then $\delta
E \delta t = p'x'dt dt = p'dt x'dt= \delta p \delta x = h$
where $h$ is Planck's constant divided by $2 \pi$.

From the other point of view one can write $\delta \nu
\delta T =1$, where $\nu$ is a frequency, $T$ is a period.
Certainly  $dT=dt$ and we see that $\delta E = h \delta
\nu$. So, one can see that $E = h \nu$ since the shift
in potential, i.e. in $E$ is possible.

\subsection{The amplitude of transition}

It is necessary to give a theory for an elementary system
which does not have fixed $x$, $v$. Since $x$ is regarded as
a main characteristic and $v$ as additional we imply that
at $t_1$
there is   $x_1$ and at $t_2$ there is $x_2$.
It is necessary to get  an amplitude of such transition.

Since the system does not have fixed $x$, $v$ one can not
speak about the fixed point in the phase space but only
about some probability to be in a fixed state in the phase
space. We have already presented a transition from the
fixed state to a distribution in a section devoted to
statistical mechanics, but it is clear that this approach
can not lead to a true result because in $\exp(-\beta H)$
the main role is played by the states with a minimal $H$.
The less is the energy the greater is the weight of the
state. In construction of the distribution for a separate
system one can not use  $H$ and has to replace it by the
characteristic of the system which has minimum on
the classical trajectory. Rigorously speaking this
characteristic $S$ has to manifest three properties
\begin{itemize}
\item
It attains minimum at the classical trajectory or somewhere
near this trajectory
\item
It has to be an additive function of two non interacting parts
of the whole system
$$
S[1,2]=S[1]+S[2]
$$
\item
It has to be an additive function of time
$$
S(t_0,t_2) = S(t_0,t_1)+S(t_1,t_2)
$$
\end{itemize}

The last two properties are necessary to consider
$\exp(\alpha S)$ as some elementary probability or
something connected with the probability. The first
property ensures the correspondence between the classical
theory and this approach.

Fortunately it is easy to present such characteristic
$$
S(t_0,t_1) =  \int_{t_0}^{t_1}(m v^2/2 - U(x)) dt
$$

The Euler equation for this functional coincides with
equation  of motion.

Then the amplitude $K$ of transition can be presented as
$$
K(a,b) = \sum_{All\ trajectories\ x(t)\ going\ from\ the\
state\
a\ to\ the\ state\ b} \exp(\alpha S(a,b))
$$

\subsection{Classical limit}

At first one has to see that the Euler equation for $S$ is
the classical equation of motion. The functional $S$ can be written
as
$$
S = \int_{t_a}^{t_b} L(x,x',t) dt
$$
where
$$
L = m v^2 /2 -U(x)
$$

Having considered a
variation of trajectory $x(t) \rightarrow x(t)+\delta x(t)$
we get
$$
S(x + \delta x) = S(x) + \int[\delta x' \frac{\partial
L}{\partial x'} +\delta x \frac{\partial
L}{\partial x} ] dt
$$
To see this equation it is necessary to prove that $x$ and
$x'$ are really independent variables but it i possible to
do already in $C^{\infty}$.

Then having integrated by parts and assuming that
$$
\delta x
\frac{\partial L}{\partial x'}|_{t_a} -
\delta x
\frac{\partial L}{\partial x'}|_{t_b} = 0
$$
which is evident since $x$ is fixed at $t_a$ and $t_b$
one can see that
$$
\delta S = - \int \delta x  [ \frac{d}{dt} (\frac{\partial
L}{\partial x'}) - \frac{\partial
L}{\partial x }]
$$
and due to the arbitrary variation   $\delta x $ one gets
$$
[ \frac{d}{dt} (\frac{\partial
L}{\partial x'}) - \frac{\partial
L}{\partial x }] =0
$$
at the minimum (maximum) of $S$.
Since $L= mv^2/2 - U$ the last equation is reduced to
$$
m x'' = - \frac{\partial U}{\partial x}
$$
which is the classical equation of motion.

Now it is necessary to investigate the limit of this
approach for macroscopic systems to show that the
trajectory goes to the classical limit.

The approximately
additive character of potential energy $U$ and kinetic
energy $\sum_i m_i v_i^2 /2$ is very important for future
analysis. Then the energy $H$ and action $S$ are the
additive functions also.

For macroscopic objects the value of $S$ is proportional
to the number of particles $N$ and, thus,
the characteristic value of relative deviation $\Delta x$ of
trajectory from providing minimum is proportional to
$N^{-1/2}$. It is very small and for macroscopic systems
trajectory is the classical one.

\subsection{Functional integral}

The calculation of the amplitude
$$
K(a,b) = \sum_{All\ trajectories\ x(t)\ going\ from\ the\
state\
a\ to\ the\ state\ b} \exp(\alpha S(a,b))
$$
is rather difficult to fulfill and can be made by the
formalism of the functional integration. Details can be
found in \cite{Fein}. For our purposes it is important that
the integral
$$
K(a,b) = \int \exp(\alpha S(a,b)) D[x]
$$
where $D[x]$ is the infinite number of differentials in
every point of trajectory
or in other notations
$$
K(a,b) = \int_{-\infty}^{\infty} \exp(\alpha S(a,b)) dx
$$
is absolutely the same expression.

The formalism of the functional integral is well
defined only in some special cases. One can restrict the
class of trajectories, for example, consider the broken
lines. This way requires the special limitations on
trajectories which are not known. But for the special
subintegral functions one has no need to make such
restrictions. Namely, for the subintegral functions of the
gaussian form
$$
f \sim \exp(Square\  form\  of\  trajectory)
$$
one can define the functional integral. The cause is the
Pyphagorean theorem which allows to squeeze the number of
differentials.

Fortunately, in the simplest cases of free dynamics (see
the section "Simple cases") the function $S$ is the square
function of coordinates  and the functional integral can be
calculated.

One can see that the principle of squeezing coordinates is
here the necessary condition to fulfill the calculations of
the functional integral. Certainly here
one has to observe a condition
$$
Re(\alpha) \leq 0
$$
if $S$ is restricted from below.

\subsection{Schr\"odinger equation}

At first one can see the principle of superposition
$$
K(b,a) = \int_{x_c} K(b,c) K(c,a) dx_c
$$
where the integration is taken over all positions of
trajectory at $t_c$.

This property allows to introduce some basic amplitudes
$K(x_i, t_i; x_0, t_0)$ giving them a name "the wave
function" $\psi(x_i,t_i)$. Then instead of
$$
K(x_2,t_2;x_0,t_0) =
\int_{-\infty}^{\infty}
K(x_2,t_2;x_1,t_1)
K(x_1,t_1;x_0,t_0)
d x_1
$$
we have
$$
\psi(x_2,t_2) =
\int_{-\infty}^{\infty}
K(x_2,t_2;x_1,t_1)
\psi(x_1,t_1)
d x_1
$$

Consider $S=\int_{t_0}^{t_1} L(x',x)dt$ for small intervals
$\epsilon = |t_1-t_0|$. Then
$$
S = \epsilon L(\frac{x-y}{\epsilon},\frac{x+y}{2})
$$
where
$x$ is initial value of trajectory and $y$ is the final
value of trajectory.
Then
$$
\psi(x,t+\epsilon) =
\int_{-\infty}^{\infty} \frac{1}{A}
\exp(\epsilon \alpha L(\frac{x-y}{\epsilon},\frac{x+y}{2})
) \psi(y,t) dy
$$
and $1/A$ is the normalizing factor.

Taking into account the explicit form of $L$ one can get
$$
\psi(x,t+\epsilon) =
\int_{-\infty}^{\infty} \frac{1}{A}
\exp( \alpha \frac{m(x-y)^2}{2\epsilon})
\exp(\epsilon \alpha U(\frac{x+y}{2},t)
) \psi(y,t) dy
$$

Having introduced $\eta = y -x$ one gets
$$
\psi(x,t+\epsilon) =
\int_{-\infty}^{\infty} \frac{1}{A}
\exp( \alpha \frac{m\eta^2}{2\epsilon})
\exp(\epsilon \alpha U(x+\eta/2,t)
) \psi(x+\eta,t) d\eta
$$

Now one can decompose $\psi$ in powers of $\epsilon$ and
get
$$
\psi(x,t) +\epsilon \frac{\partial \psi(x,t)}{\partial t} =
\int_{-\infty}^{\infty} \frac{1}{A}
\exp( \alpha \frac{m\eta^2}{2\epsilon})
(1+\alpha \epsilon U(x+\eta/2,t)
) \psi(x+\eta,t) d\eta
$$

One can see that for $$ 2 \epsilon \alpha m \sim \eta^2$$
the characteristic cancellation takes place and, thus,
$\eta$ has also some smallness. So, it is necessary to
decompose in powers of $\eta$ which gives
$$
\psi(x,t) +\epsilon \frac{\partial \psi(x,t)}{\partial t} =
\int_{-\infty}^{\infty} \frac{1}{A}
\exp( \alpha \frac{m\eta^2}{2\epsilon})
(1+\alpha \epsilon U(x+\eta/2,t)
)[\psi(x,t) + \eta \frac{\partial \psi}{\partial x}
+ \frac{1}{2} \eta^2  \frac{\partial^2 \psi}{\partial x^2} ]d\eta
$$
and
$$
\psi(x,t) +\epsilon \frac{\partial \psi(x,t)}{\partial t} =
\int_{-\infty}^{\infty} \frac{1}{A}
\exp( \alpha \frac{m\eta^2}{2\epsilon})
(1+\alpha \epsilon U(x,t)
)[\psi(x,t) + \eta \frac{\partial \psi}{\partial x}
+ \frac{1}{2} \eta^2  \frac{\partial^2 \psi}{\partial x^2} ]d\eta
$$

In the zero order
$$
\psi(x,t) =\int_{-\infty}^{\infty} \frac{1}{A}
\exp( \alpha \frac{m\eta^2}{2\epsilon}) d\eta
\psi(x,t)
$$
and then
$$
A = (\frac{-2 \pi \epsilon}{m \alpha})^{1/2}
$$

Having calculated integrals
one gets
$$
 \frac{\partial \psi}{\partial t} =
A \psi + B
\frac{\partial^2 \psi}{\partial x^2}
$$
with $|A|=|\alpha  U|$, $|B|=|1/(2\alpha m)|$
which is the Shr\"odinger equation. So, the dynamic equation
of quantum mechanics is derived.

This derivation reproduces the analysis presented in
\cite{Fein} but for an arbitrary parameter $\alpha$. Now
the task is to determine $\alpha$.

The value of $\alpha$ can
be determined by the fact that in squeezed dimensions
the classical solution and the solution of Shr\"odinger
equation must have one and the same form because there is
absolutely no information what theory has to be applied.
Classical and quantum approaches have to coincide in the
parts where nothing is known about the system.

The classical approach gives the solution in the case when
nothing is known - this solution is described in the
section "Simple solutions".
We know this solution - in classical approach
this is the superposition of
oscillations or waves. So, quantum approach has to
lead to the same waves and oscillations, at least in their functional
form.  So,
$\alpha$ has a purely imaginary magnitude
$$
\alpha =\sim i
$$

\subsection{Sense of the wave function}

Having derived the dynamic equations one has to clarify the
sense of the amplitude $K$ or at least the wave function
$\psi$. It is clear that these objects have to be connected
with a probability. So, when we have  quantum objects
with the
wave function $\psi$ in coordinate representation,
i.e. as the function of coordinate $x$, we have the
density $n(x)$ as the probability $Pdx$ to have a particle
into interval $[x,x+dx]$ as some function $F$ of $\psi$
$$
n(x) = F(\psi(x))
$$
One have to specify this function $F$ and it can be done
from some evident requirements coming from constructions of
continuous (field) models. Certainly, quantum approach has
to allow constructions of continuous models at least to
construct the field theory. Then it is necessary to
fulfill the averaging over all wave functions $\psi$.
In statistical
mechanics construction
it is also necessary to average over all states, i.e. over
all wave functions. (Here there is no necessity to consider
restrictions to occupy one and the
same energetic level for different states
providing different statistics, one can simply forget about
them.)
So, the formalism of the functional integration naturally
appears here. The function to be averaged looks like
$\exp(G)$ where $G$ is proportional to $H$, $S$ or some
other similar function. For us it is only essential that
there will be potential energy $U$ which is included
into these functions. Potential energy $U$ has to be
written via the density $n$ as
$$
U=\int_{-\infty}^{\infty} u(x) n(x) dx
$$
Then it is necessary to fulfill the functional integration
$$
\int \exp(\alpha \int_{-\infty}^{\infty} u(x) n(x) dx )
D[\psi]
$$
with some constant $\alpha$.

We do not know the real number of coordinates in $D[\psi]$.
So, the required functional integration can be fulfilled
only when the argument of $\exp$ has the gaussian form.
Then it is necessary that
$$
n(x) \sim \psi^2
$$
The value of $n$ has to be a real number which requires to
have a real number in the rhs of the previous relation
$$
n(x) \sim |\psi|^2
$$
Alternative possibility is to take $(Re \psi)^2$ which
gives a hardly appropriate result of a
quickly oscillating function.

One can see that here again the Pyphagorean theorem plays the
fundamental role and the squeezing principle works to
determine the sense of $\psi$.

Having recalled that $n$ is proportional to the probability
we see that the wave function has a simple physical
meaning:
$|\psi|^2$ is the differential probability to find a
quantum object in an elementary interval near $x$.

\subsection{Procedure of measuring}

Procedure of measuring is certainly the interaction between
the quantum object under investigation and the macroscopic
object giving the result of measuring. One has to realize
that it is impossible to give the detailed theory of such
interaction and certainly every type of measurements has
its own particular features and, thus, the detailed theory
describing the measurements. But one can come to the main
features of the measuring procedure already from the general
principles neglecting the concrete picture of interaction
between the quantum particle and the classical object.

Let us speak about the quantum particle and the measuring
device. The action of the device will be described by
an operator $A$. The elementary interaction between the
particle and the device will be presented by application of
an operator $A$ to a wave function $\psi$. Since operator $A$
is an arbitrary one, here no supposition is made. When we
speak about the additive character then it is reasonable to
take $A$ as a linear operator.
The process of interaction between the particle
and the device can not be controlled - one can not say how
many times the particle interacts with the device. So, the
final state has to be stable - the next application of $A$
does not essentially violates the wave function. Then it is
necessary that
$$
A\psi_{final} \sim \psi_{final}
$$
Then $\psi_{final}$ is an eigenfunction of $A$.

The evident physical reason of such requirement is the
observation of only resonances. It is known that the
problem of eigenfunctions appears in classical mechanics
when the resonances of oscillating systems are
investigated. From other side the resonances are the most
clear observed peculiarities of the system and there is
absolutely nothing strange that namely resonances are
considered as the observable features while all other features
are neglected.

It is known that a linear operator $A$ has many
eigenfunctions. What eigenfunction will be taken here?
Now it is necessary to decide with what intensity
(probability) the result of measuring is some $\psi_i$
$$
A_i \psi_i = a_i \psi_i
$$
where
$a_i$ is an eigenvalue.

We see that for the macroscopic flow of particles with wave
functions $\psi$ (it will be marked as $\Psi$) the result
will be $\alpha_i$ functions $\psi_i$ . Here $\alpha_i$ are some
stable characteristics.

After the procedure of measurement we have instead of
$\Psi$ the sum $\sum_i \alpha_i \psi_i$.
$$\Psi \rightarrow \sum_i \alpha_i \psi_i$$
At first there is no requirement that there is an equality
between $\Psi$ and $\sum_i \alpha_i \psi_i$. But if we make
a measurement by operator $A$ of the characteristic $a$,
then by operator $B$ of characteristic $b$,  etc, then it is
necessary that the next measurement does not feel the
previous one. For unique particle this is not true, but for
macroscopic flow this has to be observed, certainly, at the
imaginary level. We do not know who at when makes
observations. But we have to adopt that somebody very curious
makes this observation without any traces. Then we have to have to
possibility to ignore this observation.

The recipe of observation has to be one and the same for
every observation. It can not depend on the previous
observation.

To exclude the influence of the
previous observation for the macroscopic flow we need
a linear law of reconstruction $\Psi$ on a base of
$\psi_i$.
Then $\Psi$ has to be reconstructed as a
linear combination of $\psi_i$. Then
$$\Psi = \sum_i b_i \psi_i$$
and $b_i$ are coefficients in decomposition. Then
since $|\psi|^2$ is the probability one can see that
$|\alpha_i|^2$ are the probabilities to get $\Psi_i$ and
the result $a_i$ of the measurement.
Then $\alpha_i = a_i =b_i$.

As the result we see that the necessary features of
observation can be established without detailed description
of interaction. Certainly, to fulfill the requirements of
real eigenvalues and completeness the operators $A$ have to
be self-adjoint ones in a corresponding space.

\subsection{Conclusions}

In this review all constructions are based on the conception
of the measurements. At first it is introduced in classical
mechanics. Then the quantum mechanics is regarded as some
regularization of equations appeared in
classical mechanics. This way is used in other
branches of science, certainly one can consider the equations of
hydrodynamics as some very advanced way to make the
equations of motion more stable. Here the same idea is used.

All conclusions made above have to be checked many times
before they can be regarded as reliable features
of mechanics. But even now it is clear that attempts to
build a primitive
physical model for every phenomenon in nature and consider
it as the absolute truth has
an evident error in its foundation.

\end{document}